\documentclass[doublecol]{epl2} 
\newcommand {\bscco}{Bi$_2$Sr$_2$CaCu$_2$O$_{8+\delta}$}
\newcommand {\uJcm}{$\mu$J/cm$^2$}

\title{Ultrafast angle-resolved photoemission spectroscopy of quantum materials}

\author{Christopher L.\ Smallwood\inst{1} \and Robert A.\ Kaindl\inst{2} \and Alessandra Lanzara\inst{2,3}\thanks{E-mail: \email{alanzara@lbl.gov}}}
\shortauthor{C.\ L.\ Smallwood, R.\ A.\ Kaindl, and A.\ Lanzara}

\institute{                    
  \inst{1} JILA, University of Colorado and National Institute of Standards and Technology, Boulder, CO 80309, USA\\
  \inst{2} Materials Sciences Division, Lawrence Berkeley National Laboratory, Berkeley, CA 94720, USA\\
  \inst{3} Department of Physics, University of California, Berkeley, CA 94720, USA
}

\abstract{
Techniques in time- and angle-resolved photoemission spectroscopy have facilitated a number of recent advances in the study of quantum materials. We review developments in this field related to the study of incoherent nonequilibrium electron dynamics, the analysis of interactions between electrons and collective excitations, the exploration of dressed-state physics, and the illumination of unoccupied band structure. Future prospects are also discussed.
}

\begin{document}

\maketitle

\section{Introduction}

Spectroscopy, the study of interactions between light and matter, has long been a driving force in physics. Enormous progress has been achieved in this field since the invention of the laser as a source of coherent light \cite{Schawlow58,Maiman60}, which quickly led to the development of pulsed lasers and nonlinear optics \cite{Franken61,Boyd}. Today, lasers and nonlinear conversion schemes are capable of producing ultrashort pulses of durations ranging from picoseconds down to a few femtoseconds, and spanning a range of wavelengths from the terahertz to the ultraviolet \cite{Weiner}. Intense table-top lasers have been more recently employed to generate extreme-ultraviolet and X-ray pulses, enabling the creation of the shortest pulses to date, with a duration of only 67 attoseconds \cite{Zhao12}.

This impressive progress has, in turn, enabled spectroscopy of matter at the highest possible temporal resolution. In tandem with this, it is becoming recognized that some of the most interesting properties of solid-state quantum materials involve microscopic dynamics at such time scales. In the past years, numerous groups have turned the powerful eye of ultrafast spectroscopy toward the study of correlated and nanoscale quantum materials, such as superconductors, topological insulators, and charge density wave materials. Fascinating effects, which do not exist at equilibrium, are beginning to be unearthed. 

Among the most exciting developments within this field are those combining time-resolved techniques with the power of angle-resolved photoemission spectroscopy (ARPES). In equilibrium, ARPES is capable of resolving the occupied electronic structure in momentum space, uniquely revealing the dispersion of quasiparticles and many-body correlations in materials ranging from bulk and low-dimensional semiconductors, to metals, to superconductors and strongly correlated solids. By perturbing a material with an ultrashort laser pulse and following the resulting transient ARPES spectra in the time-domain, one can obtain insight into the dynamics both of quasiparticle occupations and of the electronic structure itself. In this way, important hidden physics of quantum materials can be brought to light. To this end, this perspective reviews the impact of recent advances in time-resolved ARPES on the study of quantum materials.

\begin{figure*}[tb]\centering\includegraphics[width=5.5in]{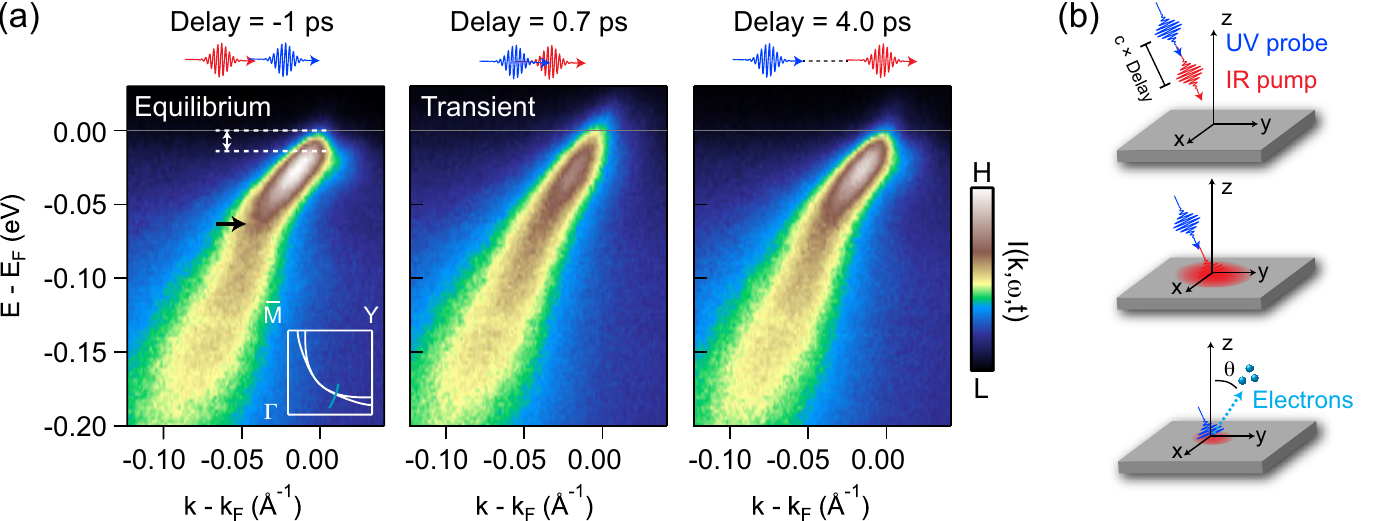}
\caption{\label{tarpes}The ultrafast ARPES technique. 
{\bf (a)} Effect of photoexcitation on the superconducting state of \bscco\ (Bi2212).
The vertical axis is energy. The horizontal axis is crystal momentum, corresponding to a slice through a gapped portion of the band structure (see Fermi surface inset). Evolution of nonequilibrium band structure can be observed by varying the time delay between pump and probe pulses, depicted here at -1 ps, 0.7 ps, and 4.0 ps.
{\bf (b)} Schematic of a typical experiment. A crystalline sample is irradiated in vacuum by infrared and ultraviolet pulses, separated by a variable time delay (top). The infrared pulse drives the material's electrons into a nonequilibrium state (center). Following this, the ultraviolet pulse photoexcites electrons into the vacuum, where their energies and momenta are measured to obtain a snapshot of band structure at time $t = \textrm{delay}$ (bottom).
}
\end{figure*}

The perspective is organized into six sections. We begin with a summary of the time-resolved ARPES technique. Following this, advances in the study of ultrafast dynamics in quantum materials are reviewed, divided into four categories:\ (a) developments where the pump pulse is used to create incoherent nonequilibrium dynamics; (b) developments where the pump pulse results in coherently oscillating collective excitations, such as phonons and charge density waves; (c) developments in the creation and monitoring of dressed-state phenomena; and (d) developments in 2-photon-photoemission (2PPE), in which pump and probe pulses are used to illuminate and explore unoccupied portions of band structure. Finally, ongoing development and future prospects in the field are discussed. 

\section{The nuts and bolts of ultrafast ARPES}

At its core, time-resolved or ultrafast ARPES consists of two wedded experimental principles: photoemission, and pump-probe spectroscopy. In a photoemission experiment, ultraviolet (UV) photons are incident on a crystalline material in vacuum. If the photon energy exceeds the the material's work function (typically about 4.5 eV), electrons are ejected into free space where their energies and exit angles can be measured. The power of the technique, particularly in its momentum-resolved form, lies in the fact that relatively simple conservation laws exist to ensure that the energy and momentum of the photoelectrons in vacuum are directly related to the energy and crystal momentum of these same electrons when they were still inside the material. As a result, the technique reveals quite direct experimental maps of electronic band structure, and has been successfully used to characterize a wide variety of materials~\cite{Hufner,Damascelli03,Hasan10,Kordyuk14,Basov14}. As an example, in the left frame of Fig.~\ref{tarpes}(a), an ARPES spectrum of the cuprate high-temperature superconductor \bscco\ (Bi2212) yields a momentum-resolved measurement of the superconducting order parameter through a directly measurable electronic excitation gap (dashed lines and white double arrow). It also displays evidence for electron-boson coupling in the form of reduced band velocity at low binding energies relative to high binding energies, and in the form of an accompanying dispersion kink at $-70$ meV (horizontal black arrow).

In spite of its advantages, equilibrium ARPES still suffers from a few drawbacks: (1) measurable electronic states must initially be occupied; (2) although important information on interactions can be gleaned from the self-energy of the dispersion, much of the microscopic dynamics cannot be deconvolved from the lineshape analysis; and (3) unique physical phases may become accessible only after driving the system far from equilibrium. These issues are circumvented by incorporating pump-probe techniques, where instead of shining only ultraviolet light on a sample, the material is irradiated in short succession by two femtosecond-duration light pulses. One of these, typically tuned to an infrared frequency and designated as the ``pump,'' drives the material system into a nonequilibrium state, but does not yet result in photoemitted electrons. The other pulse, typically designated as the ``probe,'' provides a photon energy within the vacuum ultraviolet ($\approx$ 6 eV) or extreme-ultraviolet ($\approx$ 10--60 eV) frequency range, resulting in photoemission of electrons that are ultimately detected. Nonequilibrium dynamics are studied by monitoring changes in the electron emission spectrum as the time delay between pump and probe pulses is varied. 

Figure \ref{tarpes} shows an example of the experimental data that can be acquired using the technique to measure Bi2212 [Fig.\ \ref{tarpes}(a)], as well as a schematic example of a typical ultrafast ARPES experiment [Fig.\ \ref{tarpes}(b)]. In Fig.~\ref{tarpes}(a), the left panel is acquired at negative delay time, which corresponds to an equilibrium ARPES spectrum. At a positive delay of 0.7 ps (center panel), significant changes occur in the ARPES plot, revealing both a modification of the underlying electronic dispersion as well as an overall reduction of the intensity. These changes measured in Bi2212 signify a reduction of superconductivity and a modification of the electron-boson interaction within the material, to be discussed in more detail in the sections to follow. At a delay of 4 ps, the changes are less pronounced (right panel), marking a tendency of the band structure to relax back toward equilibrium on longer timescales.

Time-resolved experimental apparatuses have been under development since the 1980s~\cite{Bokor89,Haight95,Petek97}, although the technology continues to mature with improved momentum, energy, and time resolutions, as well as increased sensitivity and tuning ranges. Notable developments include the replacement of 1D time-of-flight electron spectrometers with 2D hemispherical analyzers~\cite{Smallwood12a,Sobota12}, or even 3D time-of-flight spectrometers~\cite{Kirchmann08,Ohrwall11}, the incorporation of high-harmonic sources to generate probe photons that can access large values of electronic crystal momentum~\cite{Bauer01,Siffalovic01,Mathias07,Dakovski10}, and the incorporation of tunable pump frequencies to aid in selective nonequilibrium photoexcitation~\cite{Wang13,Parham15}.

\section{Nonequilibrium dynamics in the incoherent regime}

One of the first and still most important ways in which time-resolved spectroscopy has had an impact on correlated quantum materials research has been its ability to use near-infrared pump pulses in the incoherent regime to generate and monitor metastable nonequilibrium states. 

A scheme where the observation of these dynamics has proven particularly useful is that in which a sample is illuminated with a near-infrared pump pulse, and then measured using a probe pulse obtained using fourth-harmonic generation with $\beta$-barium borate (BBO) nonlinear crystals. In 2007, this scheme was used---with a 1.5 eV pump pulse and 6 eV probe pulse---to acquire the first truly momentum- and time-dependent ARPES data on any material, measuring quasiparticle dynamics in Bi2212~\cite{Perfetti07}. With an energy resolution of 50 meV, a momentum resolution of 0.045\ \AA$^{-1}$, and a substantial pump fluence of 100 \uJcm, the authors of this initial work found relatively little temperature dependence to their results. The findings were associated with a quasi-thermal model of metallic quasiparticle relaxation~\cite{Allen87} that was in turn used to estimate the normal-state electron-phonon coupling constant. Quasi-thermal models have also been used to derive estimates of electron-boson coupling in a number of materials beyond cuprates, including gadolinium~\cite{Bovensiepen07} and the parent compounds of iron-based superconductors~\cite{Rettig13}.

\begin{figure}[tb]\centering\includegraphics[width=80mm]{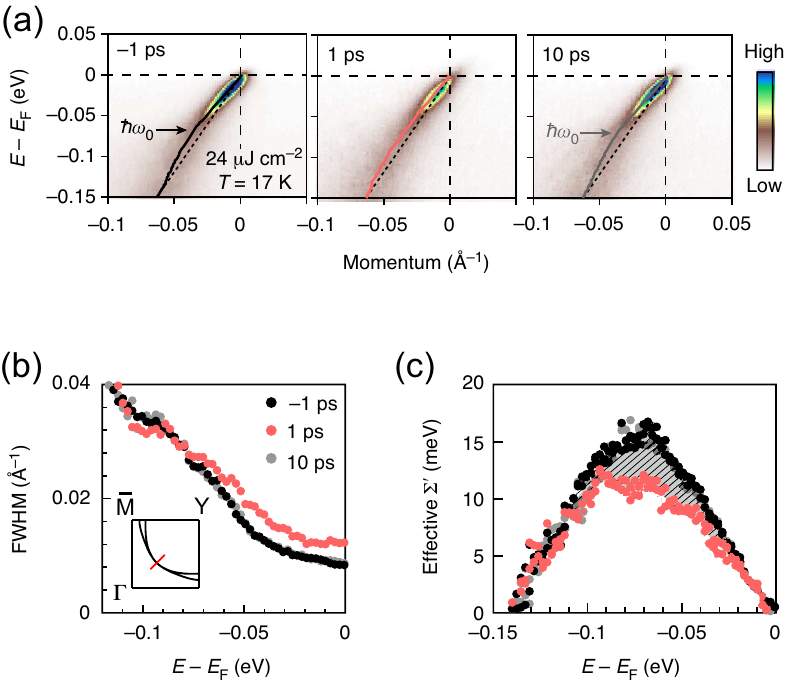}
\caption{\label{kink}Effect of photoexcitation on the 70-meV dispersion kink in \bscco\ (Bi2212), along the $\Gamma$--$Y$ direction in $k$-space (same experimental run as Fig.~\ref{tarpes}, but different momentum cut). {\bf(a)} Bosonic renormalization is visible as a kink in the equilibrium dispersion of Bi2212 (-1 ps, left). Following a near-infrared pump pulse, quasiparticle spectral weight is significantly weakened inside the kink binding energy, and the dispersion kink weakens (1 ps, center). Equilibrium features are largely recovered by 10 ps (right). 
{\bf(b)-(c)} Nonequilibrium dynamics of the self-energy $\Sigma = \Sigma' + i \Sigma''$ can be quantified by fitting the time-dependent data in (a) to Lorentzian curves, associating the fitted full-width at half-maximum (FWHM) values with an effective $\Sigma''$ [panel (b)], and subtracting a linear bare band from fitted peak positions to obtain an effective $\Sigma'$ [panel (c)]. Adapted from Zhang {\it et al.}~\cite{Zhang14}. Reprinted with permission from the Nature Publishing Group.}
\end{figure}

More recently, we and others have improved significantly on the energy and momentum resolution of these initial works. By pumping at lower fluence values than those previously explored, a number of studies using 1.5 eV pump/6 eV probe time-resolved ARPES have shown that it is possible to explore a wide variety of phenomena associated not just with metallic-state dynamics, but also with dynamics associated with the superconducting ground state of cuprates~\cite{Graf11,Cortes11,Smallwood12,Zhang13,Rameau14,Smallwood14,Zhang14,Miller15,Yang15,Piovera15,Smallwood15,Ishida16,Smallwood16,Zhang16}. Evidence for the electron-boson coupling interaction within this low-fluence, low-temperature regime can be observed in time-resolved ARPES spectra of cuprate superconductors through a variety of features. Among the most striking is the dramatic effect that a 1.5 eV pump pulse has on a well-known 70-meV kink that appears in the dispersion of cuprate superconductors along the nodal direction~\cite{Lanzara01}, resulting from electron-boson coupling. As shown by Fig.~\ref{kink}, photoexcitation at 1.5 eV in Bi2212 results in a suppression in quasiparticle spectral weight in the nodal ARPES spectra [Fig.~\ref{kink}(a), middle panel]~\cite{Graf11}, as well as an increase in ARPES linewidth [Fig.~\ref{kink}(b)]~\cite{Zhang14,Ishida16}, which is confined to binding energies less than 70 meV. In addition, the pump induces a marked change in band velocity~\cite{Zhang14,Ishida16}. This is particularly visible in Fig.~\ref{kink}(c), where a linear bare band approximation has been subtracted from fitted dispersions in Fig.~\ref{kink}(a) to generate an effective real part of the electronic self-energy. Theoretical work \cite{Sentef13,Kemper14} has demonstrated that a weakened dispersion kink does not necessarily imply a fundamental reduction in electron-boson coupling. However, it is striking to note that kink effects become sharply reduced when the equilibrium temperature is increased above $T_c$ (though some effects are still evident at the highest pump fluences~\cite{Rameau14}), which may indicate a connection to the mechanism of superconductivity.

\begin{figure}[tb]\centering\includegraphics[width=80mm]{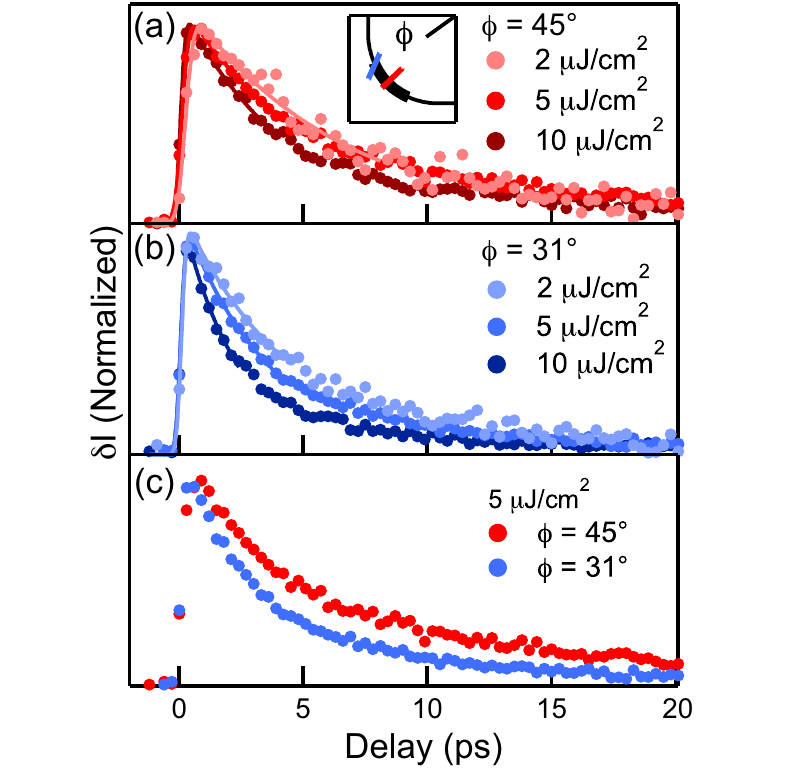}
\caption{\label{fluence}Fluence- and momentum-dependent quasiparticle relaxation in the superconducting state of Bi2212. {\bf(a)-(b)} Quasiparticle relaxation rates, integrated within a range of energy and momenta along nodal (a) and off-nodal (b) slices through the Brillouin zone, are fluence-dependent, with faster recovery occurring at higher fluences. {\bf(c)} At fixed fluence, quasiparticle relaxation rates are faster away from the node than at the node. Adapted from Smallwood {\it et al.}~\cite{Smallwood12}. Reprinted with permission from AAAS.}
\end{figure}

Beyond its ability to resolve changing band velocities and bosonic renormalization effects, time-resolved ARPES has opened windows into the interactions between quasiparticles and the superconducting condensate that have no analogue in equilibrium ARPES. Recently, for example, studies have measured the dynamics of the excitation gap in Bi2212 at both low and intermediate fluence~\cite{Smallwood12,Smallwood14,Zhang14}, establishing a measurement of the critical fluence necessary to destroy superconductivity, and characterizing the timescales on which these processes occur. Perhaps more interestingly, in the presence of a gap, quasiparticle relaxation rates in the cuprates become markedly both fluence-dependent~\cite{Gedik04,Kaindl05,Gedik05,Smallwood12,Yang15,Smallwood15} and momentum-dependent~\cite{Smallwood12}. As shown by Fig.~\ref{fluence}, fluence-dependent quasiparticle dynamics have been observed along a variety of momentum directions in Bi2212, both along a cut intersecting the superconducting gap node [Fig.~\ref{fluence}(a)] and in regions of $k$-space where the superconducting gap is finite [Fig.~\ref{fluence}(b)]. Moreover, at constant pump fluence, the quasiparticle relaxation rate increases markedly with increasing gap size [Fig.~\ref{fluence}(c)]. The results open possibilities for using the momentum-dependent structure of relaxation rates to access momentum-dependent coupling effects between superconductivity and charge density waves, antiferromagnetic order, and phonons.

Though time-resolved ARPES studies making use of fourth-harmonic generation are unparalleled in their momentum and energy resolution, they suffer from an inability to access the full Brillouin zone of many materials. Recent years have seen increasing efforts to combine ultrafast ARPES with table-top laser sources of extreme-UV light around 10--50 eV, which can be obtained by high-harmonic generation (HHG) in gases driven by intense laser pulses. The energy resolution of such HHG-based techniques is often limited by the harmonic width or monochromator settings to 150--500 meV (compared to 20 meV or less for nonlinear crystal experiments) \cite{Siffalovic01,Mathias07,Dakovski10}. Recent efforts in UV-driven XUV sources are enabling much narrower bandwidths of 70 meV directly from the source, obtained at high repetition rates of 50-kHz and with significant flux \cite{Wang15}. Ultrafast ARPES experiments making use of such techniques have begun to explore nonequilibrium dynamics in a variety of materials~\cite{Dakovski11,Gierz13,Johannsen13,Gilbertson14,Ulstrup14,Dakovski15}, with among the most prominent developments being an elucidation of the physics of charge-density-wave (CDW) materials including 1$T$-TaS$_2$, 1$T$-TiSe$_2$, and La$_{1.75}$Sr$_{0.25}$NiO$_4$~\cite{Rohwer11,Petersen11,Hellmann12b}. 

For example, developments in time-resolved ARPES have given researchers direct access to timescales for charge-order collapse, manifest experimentally through the changing magnitude of the CDW gap. Remarkably, early measurements on 1$T$-TiSe$_2$ revealed that this quenching can occur as rapidly as within 20 fs~\cite{Rohwer11}. This is in sharp contrast to the timescale for the collapse of the superconducting gap in Bi2212, which is delayed by as much as 600 fs~\cite{Smallwood14}. As high-harmonic techniques have improved, researchers have been able to compare time scales between various charge-density-wave materials, and to use this information to differentiate among excitonic, Peierls, and Mott effects driving CDW formation~\cite{Hellmann12b}. 

\section{Collective mode oscillations}

With impulsive excitation, ultrafast ARPES has also been shown capable of inducing oscillating spectral signatures reflecting the coupling of electrons to collective modes. One of the more intriguing phenomena is the observation of a pump-induced oscillation in the electronic band structure~\cite{Rettig12,Rettig13,Avigo13}, which lasts as long as 2 ps in the iron-based superconductor BaFe$_{1.85}$Co$_{0.15}$As$_2$ and in its antiferromagnetic relatives BaFe$_{2}$As$_2$ and EuFe$_2$As$_2$. The oscillations were associated with electron-phonon coupling.

Similar coherent phonon oscillations have been observed in Bi2212~\cite{Yang16}, where the oscillation frequency has been shown to be intriguingly close to an electron-phonon renormalization kink in the band structure visible also through equilibrium ARPES~\cite{Rameau09,Plumb10,Vishik10}. Interestingly, the ability to observe electron-boson coupling interactions through ultrafast coherent oscillations is complementary, and in certain respects even superior to efforts to extract electron-boson coupling interactions by looking for band renormalization effects using equilibrium ARPES. The reason is that in materials exhibiting significant three-dimensional band dispersion (like many of the iron-based superconductors), band positions become obscured by the absence of out-of-plane momentum conservation in an ARPES experiment. Even in cases where band renormalization effects are directly visible using equilibrium ARPES---as they are in the cuprate superconductors---the ultrafast determination of coupled bosonic mode energies can be much more sharply defined under ultrafast techniques than they can under equilibrium techniques. Moreover, only specific modes are excited, with coupling to different bands that can be mapped by studying the transient modulation of the band structure across momentum space.

\begin{figure}[tb]\centering\includegraphics[width=80mm]{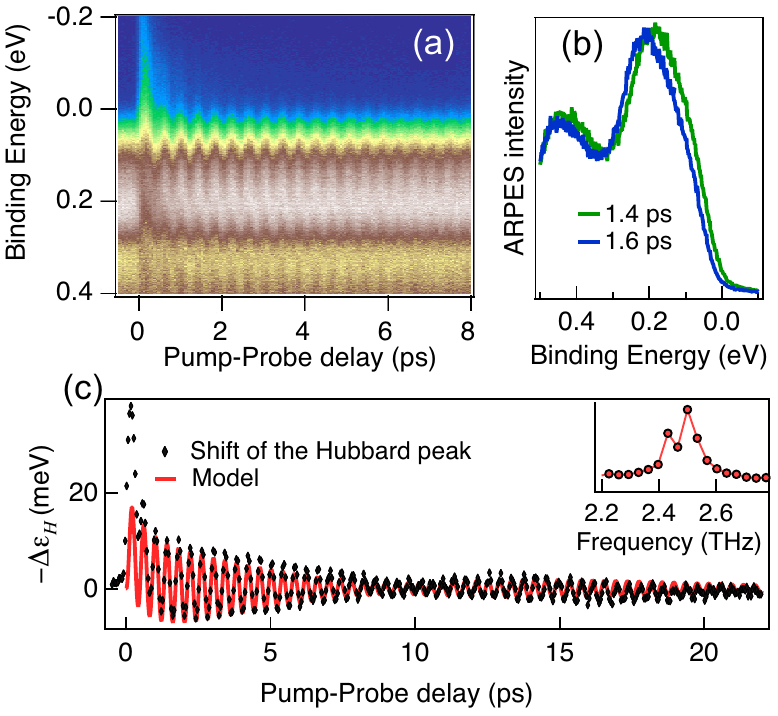}
\caption{\label{PerfettiCDW}Pump-induced gap magnitude oscillations in the CDW material 1$T$-TaS$_2$. {\bf(a)--(b)} Time- and energy-dependent photoemission spectral weight, integrated across a window in momentum, following a near-infrared (1.5 eV) pump pulse. {\bf(c)} Shift in the Hubbard peak, defined as the change in the peak energy of the photoemission signal, as a function of pump-probe delay. Inset shows the result of applying a Fourier transform to the data. From Perfetti {\it et al.}~\cite{Perfetti06}. Reprinted with permission from the American Physical Society.}
\end{figure}

In CDW systems, coherently driven oscillations appear as well. Early measurements of these oscillations in KMO by Demsar {\it et al}.\ using all-optical techniques~\cite{Demsar99} were replicated in a wide variety of materials, ranging from 1$T$-TaS$_2$~\cite{Perfetti06,Petersen11} to TbTe$_3$~\cite{Schmitt08}. One of the more prominent examples of this type of oscillation in an ultrafast ARPES spectrum is displayed in Fig.~\ref{PerfettiCDW}, where it was demonstrated by Perfetti {\it et al}.~\cite{Perfetti06} that a 1.5 eV pump pulse can induce gap oscillations in the CDW material 1$T$-TaS$_2$ lasting as long at 20 ps. 

 \section{Dressed-state physics}
 
The strong light-matter interaction between the electronic wavefunctions of a crystalline solid and the coherent electric field of an intense laser pulse can also result in novel modifications of the electronic structure. Among the most intriguing developments has been the observation of Floquet-Bloch states in topological insulators. Such states are a time-dependent extension of Bloch waves: Bloch waves are periodic in crystal momentum and result from a Hamiltonian that commutes with discrete translations in space, whereas Floquet-Bloch states are periodic in \emph{both} crystal momentum and energy, and result from a Hamiltonian that commutes with discrete translations in space and time.

\begin{figure}[tb]\centering\includegraphics[width=80mm]{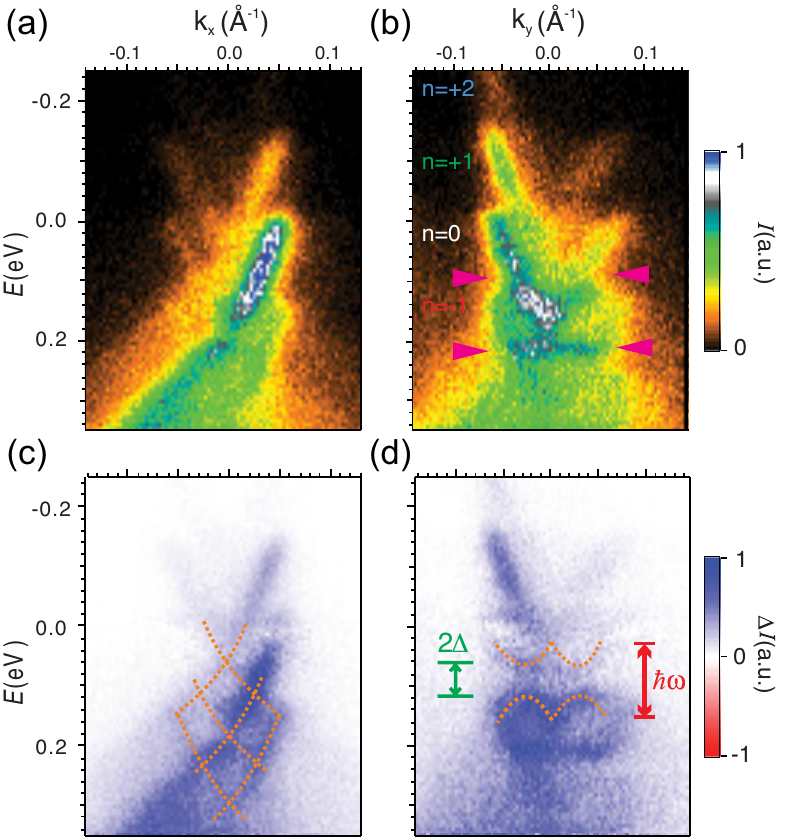}
\caption{\label{floquetbloch}Floquet-Bloch states in Bi$_2$Se$_3$. {\bf(a)} At a pump-probe delay of 0 fs, a number of replica bands appear in the time-resolved ARPES spectrum of Bi$_2$Se$_3$. {\bf(b)} Along the $k_y$-direction, which is perpendicular to the direction of the applied electric field, the intersections between these replica bands exhibit avoided crossings, which is a hallmark signature of Floquet-Bloch phenomena. {\bf(c)--(d)} Difference spectra, where the equilibrium spectra at $t=-500$ fs have been subtracted from the spectra displayed in (a) and (b). Adapted from Wang {\it et al.}~\cite{Wang13}. Reprinted with permission from AAAS.}
\end{figure}

Recently, Wang {\it et al.}\ demonstrated experimental evidence of this effect~\cite{Wang13}, as shown in Fig.~\ref{floquetbloch}. Researchers performed pump-probe measurements on the topological insulator Bi$_2$Se$_3$, and found that if the frequency of the pump pulse was reduced below the bulk band gap, and the timing of the probe pulse was tuned to within the duration of the pump pulse, then a ladder of replica bands begins to appear with evenly spaced energies in the photoemission spectrum. Most interestingly, the surface state bands in Bi$_2$Se$_3$ are dispersive, leading to a number of points where bands of different energy orders overlap. For appropriately polarized light, band gaps open up at these specialized points, with the induced gap magnitude being proportional to the strength of the applied electric field~\cite{Mahmood16}. This property separates Floquet-Bloch phenomena from more commonly observed laser-assisted photoemission (LAPE)~\cite{Miaja06,Saathoff08,Rohwer11}, which is the result of photon-dressed electron states in vacuum. The effect may facilitate device applications with ultrafast switching times.

Floquet-Bloch phenomena may also result in fascinating new phenomena in other types of materials. Calculations indicate that for sufficiently high pump pulse energies, Floquet-Bloch phenomena in the form of a pump-induced band gap can be induced in graphene~\cite{Sentef15}. For pulse energies approaching very high values, the gap may even begin to shrink again, leading to the possibility of interesting band gap tuning depending on pump fluence.

\section{Revealing unoccupied band structure}

Beyond its ability to create both novel nonequilibrium states in both the coherent and incoherent regime, ultrafast ARPES has opened up a number of possibilities for mapping out the unoccupied band structure in solids, using 2-photon photoemission (2PPE) techniques. The ability to access these unoccupied bands is important, for example, to develop a better understanding of the conduction bands of nanoscale materials for applications in opto-electronics, or of the energy-momentum dispersion of correlated quasiparticles in quantum materials. Strictly speaking, a time-delayed pump and probe pulse is not necessarily required to reveal unoccupied-state data. If the pulse energy is sufficiently intense, the same pulse can act both as pump and probe. Still, the ability to isolate pump and probe signals using a variable delay can help isolate the signal of interest from background noise.

\begin{figure}[tb]\centering\includegraphics[width=80mm]{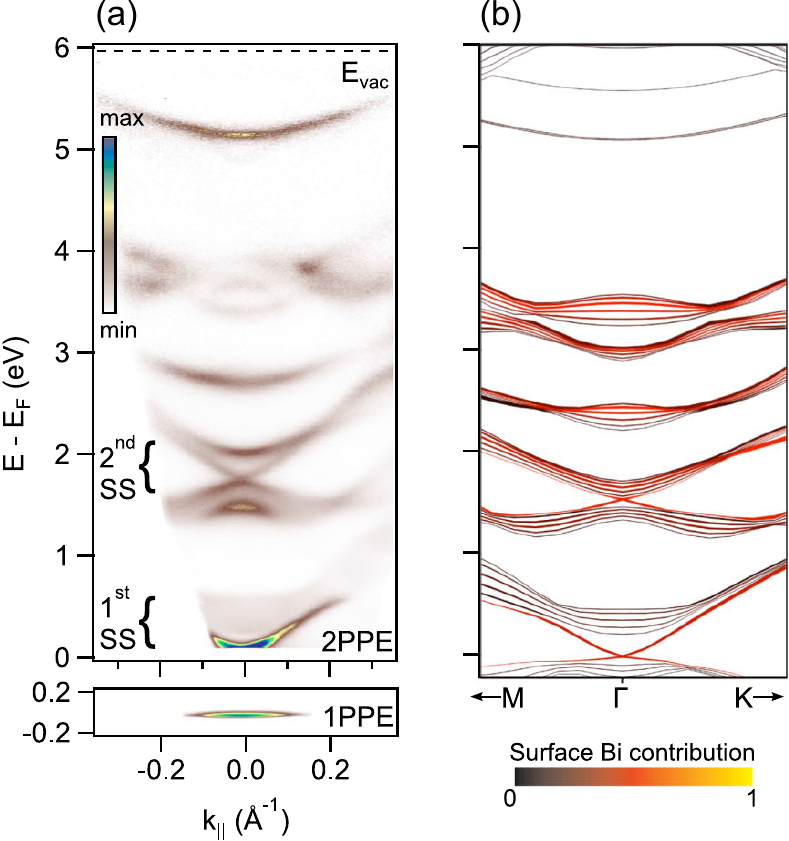}
\caption{\label{2ppe}Unoccupied band structure of $p$-type Bi$_2$Se$_3$. {\bf(a)} Traditional ARPES (1PPE) is limited to occupied states (bottom), but 2PPE (top) reveals unoccupied band structure several eV beyond the Fermi level, including Dirac cones in the unoccupied band structure that had not previously been observed. {\bf(b)} Slab band structure calculation, which agrees well with experiment. Adapted from Sobota {\it et al.}~\cite{Sobota13}. Reprinted with permission from the American Physical Society.}
\end{figure}

In topological insulators, ultrafast ARPES measurements of the unoccupied band structure are particularly striking. It has been shown that it is possible, for example, to produce maps of the unoccupied surface-state band structure in Bi$_2$Se$_3$ extending even 6 eV above the chemical potential, as demonstrated by Sobota {\it et al}.~\cite{Sobota13} (see Fig.~\ref{2ppe}). More recently, Grubi\v{s}i\'c \v{C}abo {\it et al}.\ have shown that it is possible to use ultrafast ARPES to measure the conduction band of the transition metal dichalcogenide MoS$_2$~\cite{Cabo15}, and Ulstrup {\it et al}.\ have extended these studies to a MoS$_2$/graphene heterostructure~\cite{Ulstrup16}, adding new insight into the free carrier dispersion and band gap.

\section{The future}

As ultrafast ARPES continues to mature, completely new applications and directions are beginning to emerge. Among the most ambitious and exciting new directions include possibilities for optically enhancing superconductivity in high-$T_c$ materials, tailoring pump light in increasingly sophisticated ways to achieve coherent control of quantum phases, and developing the ability to acquire simultaneous time-resolved, momentum-resolved, and spin-resolved ARPES data.

Tantalizing possibilities for ultrafast ARPES have to do with the prospect of using optical means to access ``hidden'' states of materials, which are thermodynamically inaccessible, but which can be accessed using intense optical pulses. Two examples, which have already been explored using all-optical techniques, are the possibility of generating a metastable conducting state in the CDW material 1$T$-TaS$_2$~\cite{Stojchevska14}, and that of generating photoinduced superconductivity at temperatures above the equilibrium critical temperature in cuprate superconductors~\cite{Fausti11,Hu14,Kaiser14}.

Beyond this, there are intriguing nonlinear technical possibilities on the horizon for ultrafast ARPES. A particular example is the prospect of incorporating increasingly sophisticated pump pulse sequences in ARPES experiments. At present, nearly all operational ultrafast ARPES experiments can be understood in the regime of ``linear'' response: the pump pulse is understood to transfer quasiparticles from lower energies to higher energies in a manner that can be modeled using Fermi's golden rule. Even in cases where coherent oscillations are generated, the use of this single pulse has an only limited ability to control these oscillations. In contrast, all-optical techniques have demonstrated that a great deal of increased control can be achieved if the pump pulse is divided into multiple sub-pulses with variable time delays between pulses. One initial proof-of-principle demonstration of using these techniques was demonstrated by Aeschlimann {\it et al}. in 2011~\cite{Aeschlimann11}.

Finally, recent advances in spin-resolved ARPES~\cite{Jozwiak10,Gotlieb13} are opening up possibilities to establish an experiment capable of simultaneous time, momentum, energy, and spin resolution in crystalline materials. This capability should prove useful in the generation and analysis of nonequilibrium states in materials with spin-textured band structure. Topological insulators are an obvious example.

As we have outlined in this perspective, ultrafast ARPES has in recent years evolved into a powerful technique enabling direct access to the dynamics of quasiparticles, electronic structure, and collective modes of nanoscale and correlated quantum materials. While the insight gained is already remarkable, we can expect the rapid pace of progress to continue, portending insights into the properties of solids in time, energy, and momentum space that remain hidden in equilibrium.

\acknowledgments
This work was supported as part of the Ultrafast Materials Program at Lawrence Berkeley National Laboratory, funded by the U.S.\ Department of Energy, Office of Science, Office of Basic Energy Sciences, Materials Sciences and Engineering Division, under Contract No.\ DE-AC02-05CH11231\@. C.L.S.\ acknowledges support from an NRC Research Associateship award at NIST.


\end{document}